\documentclass[a4paper,conference]{IEEEtran}
\usepackage{indentfirst}
\usepackage{booktabs}
\usepackage{enumitem,amsmath,amssymb}
\usepackage{bm}
\usepackage{array}
\usepackage{multicol}
\usepackage{multirow}
\usepackage{graphicx}
\usepackage{color}
\usepackage{url}
\ifCLASSOPTIONcompsoc
 \usepackage[caption=false,font=normalsize,labelfont=sf,textfont=footnotesize]{subfig}
\else
 \usepackage[caption=false,font=footnotesize]{subfig}
\fi

\begin{document}
\title{Multimodal Speech Enhancement Using Burstpropagation}
\author{
    \IEEEauthorblockN{Mohsin Raza$^1$, Leandro A. Passos$^3$, Ahmed Khubaib$^1$,  and Ahsan Adeel$^3$\\}
    \IEEEauthorblockA{1. University of Wolverhampton, Wolverhampton, England, UK\\
    2. São Paulo State University, Bauru, SP, Brazil\\
    3. Stirling University, Stirling, Scotland, UK\\
    M.R.Naseem@wlv.ac.uk} }
\maketitle

\begin{abstract}
This paper proposes the MBURST, a novel multimodal solution for audio-visual speech enhancements that consider the most recent neurological discoveries regarding pyramidal cells of the prefrontal cortex and other brain regions. The so-called burstpropagation implements several criteria to address the credit assignment problem in a more biologically plausible manner: steering the sign and magnitude of plasticity through feedback, multiplexing the feedback and feedforward information across layers through different weight connections, approximating feedback and feedforward connections, and linearizing the feedback signals. MBURST benefits from such capabilities to learn correlations between the noisy signal and the visual stimuli, thus attributing meaning to the speech by amplifying relevant information and suppressing noise. Experiments conducted over a Grid Corpus and CHiME3-based dataset show that MBURST can reproduce similar mask reconstructions to the multimodal backpropagation-based baseline while demonstrating outstanding energy efficiency management, reducing the neuron firing rates to values up to \textbf{$70\%$} lower. Such a feature implies more sustainable implementations, suitable and desirable for hearing aids or any other similar embedded systems.

\end{abstract}


\section{Introduction}
\label{s.intro}


The World Health Organization states that $430$ million people suffer from moderate to higher hearing loss nowadays and estimates that almost $2.5$ billion will have hearing impairment to some degree by 2050~\cite{world2021hearing}. The problem impacts individual social relationships and the perception of surrounding sounds~\cite{noble1998self}, which may also lead to psychological distresses~\cite{huang2021hearing,helvik2006psychological}. Therefore, intelligent computer systems that clean the audio signal are highly desirable.

In this context, many efforts have been applied toward machine learning-based approaches for speech enhancement~\cite{das2021fundamentals,rehr2017importance}. A particularly interesting approach comprises multimodal approaches that combine correlated audio and visual (AV) information to amplify relevant information and suppress noise~\cite{ngiam2011multimodal}. The idea can be illustrated by a dialogue in a pub where people talk and live music is played in the background, and the listener focuses on reading the speaker's lips and expressions to mind the context and infer meaning from the conversation. The work of Adeel et al.~\cite{adeel2018real} comprises related applications regarding IoT and 5G for lip-reading hearing aids, encrypted audio speech reconstruction~\cite{adeel2020novel}, and speech enhancement in different conditions using deep learning~\cite{adeel2020contextual,gogate2020cochleanet}. Furthermore, recent work by Passos et al.~\cite{passos22:informationFusion} investigates the fusion of audio and visual information using graph neural networks and canonical correlation analysis, as well as a cortical cell-inspired model that approximates the computational model to a more biologically plausible approach~\cite{passos2022canonical}.

Despite such approximation, the methods mentioned above and most of the machine and deep learning solutions rely on backpropagation, an algorithm inspired by an antiquate modeling of neuronal information flow where inputs are linearly combined and exposed to an activation function, whose output feeds consecutive layers. Further, the outcome is compared to an expected target, and the error is back-propagated, assigning the credit for any mistakes or successes to neurons that are multiple synapses away from the output and updating the weights associated with such neurons accordingly. In this scenario, Payeur~\cite{payeur2021burst} proposed a novel approach inspired by more recent studies on the physiological mechanism of pyramidal neurons, where the learning is regulated by the frequency of bursts, the so-called Burstpropagation.

Burstpropagation tackles the credit assignment problem in a more biologically plausible manner by addressing the primary principles of pyramidal neurons suggested by K\"{o}rding and K\"{o}nig~\cite{kording2001supervised} as follows. First, it employs a burst-dependent learning rule to steer the sign and magnitude of plasticity through feedback stimuli. Second, it multiplexes the feedback and feedforward information across multiple layers using different connections, i.e., distinct weights are used during the forward and backward propagation processes. Third, it performs the alignment between feedback and feedforward connections by approximating the loss-function gradients through burst-dependent learning. Finally, the fourth principle regards linearizing the feedback signals concerning the credit information, which can be performed using recurrent connections.

Therefore, this paper proposes the so-called MBURST, a multimodal approach that combines audio and visual information to enhance speech quality using Burstpropagation. The model preserves the biological properties of real neurons like short-term synaptic plasticity~\cite{markram1998differential}, dendritic excitability~\cite{larkum1999new}, synaptic transmission, inhibitory microcircuit, and burst-dependent synaptic plasticity~\cite{payeur2021burst}. Experiments conducted over a Grid Corpus and CHiME3-based dataset for clean audio mask reconstruction show that MBURST can generate clean audio masks with similar quality to a multimodal backpropagation-based baseline while dramatically reducing the energy consumption, reaching values up to $70\%$ lower. Such an attribute is extremely desirable for real-world applications like embedded systems on hearing aid devices. Thus, this paper comprises two main contributions:

\begin{itemize}
    \item it provides the MBURST, a biologically plausible and energy-efficient method that combines audio and visual information for speech enhancement; and
    \item it introduces Burstpropagation, a state-of-the-art algorithm inspired by pyramidal neurons, in the context of multimodal learning.
\end{itemize}

The remainder of this paper is presented as follows. Section~\ref{s.theoretical} provides a theoretical background regarding Burstpropagation and the proposed approach. Further, Sections~\ref{s.methodology} and~\ref{s.experiments} describe the methodology employed to conduct the experiments and the results, respectively. Finally, Section~\ref{s.conclusions} states the conclusions and future work.

\section{Theoretical Background}
\label{s.theoretical}

This section briefly introduces the concepts behind Burstpropagation, as well as the proposed Burstpropagation-based multimodal approach, namely MBURST.

\subsection{Burstpropagation}
\label{ss.burstprop}

The burst-dependent algorithm, or Burstpropagation, defines the weighted sum at the neuron's input as ``somatic potentials''. Similarly, the neuron's output is termed the event rate. The 'somatic potentials' of a dense layer are defined as:
\begin{equation}
    v_{l} = W_{l}e_{l-1},
    \label{eq.somaticPotential}
\end{equation}
where $l\in\{1,2,\cdots,L\}$ denotes the network's layers, $W_{l}$ is the weight connecting layer $l-1$ to layer $l$ and $e_{l}$ is the $l$th layer's event rate, defined by:

\begin{equation}
    e_{l} = f(v_{l}),
    \label{eq.eventRate}
\end{equation}
where $f_{l}$ stands for any activation function, e.g., the Rectifier Linear Unit (ReLU) or Sigmoid, for layer $l$. Regarding convolution layers, the multiplications are simply replaced by a convolution operator where $W_{l}$ represents layer $l$'s kernel. 

Since the burst rate drives the model's learning procedure, the first step in the process comprises computing the output layer ($l=L$) bursting probability:
\begin{equation}
    p_{L} = \zeta(\Bar{p}_{L} - h(e_{L}) \odot \Delta_{e_L} \mathcal{L})
    \label{eq.burstprobTop}
\end{equation}
where $\Bar{p}_{L}$ is a hyperparameter that defines the reference bursting probability, $\zeta$ is a squashing function that guarantees $p_{L,i}\in\left[0,1\right]$, $\Delta_{e_{L}} \mathcal{L}$ stands dor the loss function derivative, and $h(e_{l})$ is a vector-valued function defined by:

\begin{equation}
    h(e_{l}) \equiv f^{'}(v_{l}) \odot e_{l}^{-1}.
    \label{eq.h_e}
\end{equation}

The bursting probability is used to calculate the bursting rate with and without teaching, i.e.,  $b_{l}$ and $\Bar{b}_{l}$, respectively, at any layer $l$:

\begin{equation}
\begin{array}{c}
        b_l = p_l \odot e_l,\\ 
        \Bar{b}_{l} = \Bar{p}_{l} \odot e_l.
\end{array}
\label{eq.burstRateTop}
\end{equation}

Further, the bursting rates are propagated to the previous layers through feedback weights, namely $Y$. The result is employed to compute the current hidden layer's ``dendritic potential'' with ($u_{l}$) and without ($\Bar{u}_{l}$) teacher. Notice that, for convolutional layers, $Y$ stands for a kernel, and a convolution operation replaces the multiplication. 

\begin{equation}
\begin{array}{c}
        u_{l} = h(e_{l}) \odot (Y_{l}b_{l+1}),\\ 
        \Bar{u}_{l} = h(e_{l}) \odot (Y_{l}\Bar{b}_{l+1}).
\end{array}
\label{eq.dendriticPotential}
\end{equation}

In the sequence, the somatic potentials are used to calculate the hidden layer's bursting probability and the reference bursting probability:

\begin{equation}
\begin{array}{c}
        p_{l} = \sigma(\beta u_{l} + \alpha),\\ 
        \Bar{p}_{l} = \sigma(\beta \Bar{u}_{l} + \alpha),
\end{array}
\label{eq.burstprobHidden}
\end{equation}
where $\beta=1$ and $\alpha=0$ are are constants that control the dendritic transfer function and $\sigma$ is a sigmoid function.

Finally, the change in the forward and feedback weights, $W$ and $Y$, respectively, are computed as follows: 

\begin{equation}
\begin{array}{c}
        \Delta W_l = \Delta Y_l = -(b_{l} - \Bar{b}_{l}) \odot e_{l-1}^{T}.
\end{array}
\label{eq.weightsChange}
\end{equation}

Notice that, once again, the multiplication is replaced by a backward convolution in the context of convolutional layers.

\subsection{MBURST}
\label{ss.proposed}

The proposed multimodal Burstpropagation-based approach to speech enhancement combines audio and visual information for clean audio estimation. The model takes advantage of recent neurological research on pyramidal cells to learn coherent relationships between noisy audio and visual information and extract clean information by amplifying relevant information and suppressing noise.

\begin{figure*}[!h]
  \centerline{
    \begin{tabular}{c}
	\includegraphics[width=\textwidth]{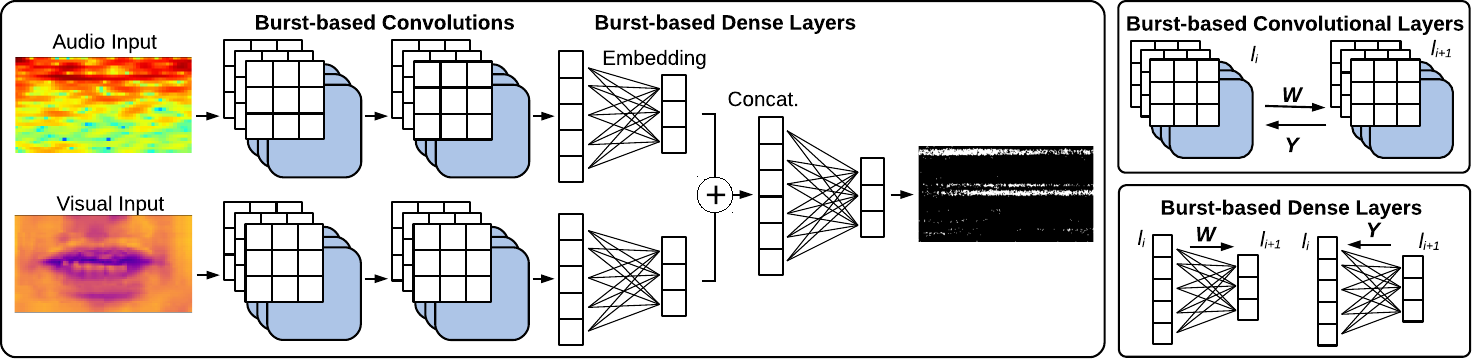} 
    \end{tabular}}
    \caption{Proposed method. The left image depicts the multimodal network with noisy audio and visual inputs. Both channels comprise two layers of burst-based convolutions, followed by flattening and a burst-based dense layer for feature embedding. Finally, such embeddings are concatenated and used to feed another burst-based dense for classification. Right-top and right-bottom frames illustrate the burst-based convolutional and dense layers, respectively. Notice such layers comprise a forward weight matrix (kernel) $\bm{W}$ and a backward weight matrix $\bm{Y}$.}
  \label{f.model}
\end{figure*}

The model comprises two burst-dependent convolutional neural networks that work in parallel to extract features from noisy audio inputs and their respective frames. The outcomes from both channels are flattened and exposed to a burst-dependent fully-connected layer for embedding extraction. The embeddings are concatenated into a single feature vector, which is used to feed a similar burst-dependent dense layer for classification. Figure~\ref{f.model} depicts the process pipeline.

\section{Methodology}
\label{s.methodology}

This section describes the dataset and the environment configuration considered in the experiments. 

\subsection{Dataset}
\label{ss.dataset}

The experiments performed in this work were conducted over a dataset based on roughly $1,000$ sentences composed of a six-word sequence extracted from Speakers $1$ in Grid Corpus~\cite{cooke2006audio} dataset. The clear Grid utterances are blended with non-stationary random noises, e.g., bus, cafe, street, and pedestrian, using four distinct Signal-to-noise ratios (SNR), i.e., $\{-12dB, -6dB, 0dB, 6dB\}$, extracted from the CHiME3\cite{barker2017third}.

\subsubsection{Data pre-processing}
\label{sss.preprocessing}

The original audio was resampled at $16$ kHz. Each utterance was divided into approximately $75$ frames with $1,244$ samples per frame and a $25\%$ increment rate. Further, a $622$-bin power spectrum was computed using STFTs and a hamming window procedure. A Viola-Jones lip detector~\cite{viola2001rapid} and an object tracker~\cite{ross2008incremental} extract the speakers' lip images from the GRID Corpus films, recorded at $25$ frames per second. A $92\times50$-inch area around the lip's center was picked. The recovered lip sequences were up-sampled three times to match the $75$ STFT frames of the audio signal. 

\subsubsection{Ideal Binary Mask}
\label{sss.binaryMasks}

Hu and Wang~\cite{hu2004monaural} proposed the so-called Ideal Binary Mask (IBM), a binary mask with a positive value to a time-frequency (TF) unity whose goal energy is greater than the interference energy and zero otherwise. Among a number of the desired qualities in speech enhancement, the binary masks retain the speech-dominated units while zeroing out the noise-dominated units of noisy speech and providing high intelligibility scores on speech reconstruction, even considering meager signal-to-noise ratio (SNR) settings~\cite{jiang2011performance}.

IBM employs a premixing of target and interference signals, i.e., the speech and noise signals, computed through a criterion

\begin{eqnarray}
    IBM(t,f) & = & \left\{ \begin{array}{ll}
        1 & \mbox{if } 10\log_{10}\left(\frac{X(t,f)^2}{N(t,f)^2}\right)\ge LC, \\
        0 & \mbox{otherwise}
        \end{array}\right. 
    \label{e.IBM}
\end{eqnarray}
where $X$ stands for the speech with no background noise, $N$ represents the noise portion, and $LC$ is the local criterion used to differentiate between speech and background noise, i.e., a threshold. Further, the frame of reference $t$ and the frequency range $f$ denote the time and frequency dimensions, respectively. 

The method assigns a positive value to the more prevalent TF units if their SNR is greater than or equal to the LC. Conversely, TF units dominated by noise are presumed to have SNRs lower than LC. Thus they are given assigned a value of 0 and muted. The proportion of the TF-positive speech versus noise-dominant samplings relies on a proper LC selection. Since SNRs greater than $0$ dB are speech dominating and those less than $0$ dB are noise dominant, an LC of 0 dB is supposedly optimal. However, studies suggest that such selection may imply disproportionate removal of TF units on mixed noisy and speech instances due to the addition of noise known as artifacts, therefore suggesting employing an LC of $5$ dB~\cite{healy2015algorithm, healy2015algorithm} for optimum performance.

\begin{figure*}[!h]
  \centerline{
    \begin{tabular}{cccc}
	\includegraphics[width=0.235\textwidth]{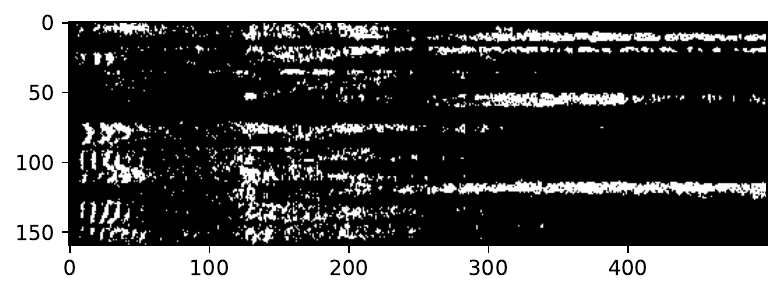} & 
	\includegraphics[width=0.235\textwidth]{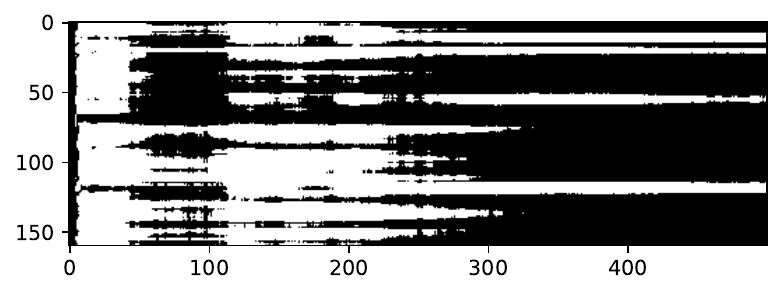} & 
	\includegraphics[width=0.235\textwidth]{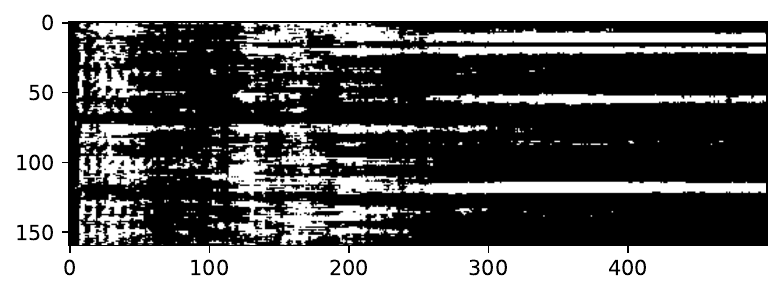} & 
	\includegraphics[width=0.235\textwidth]{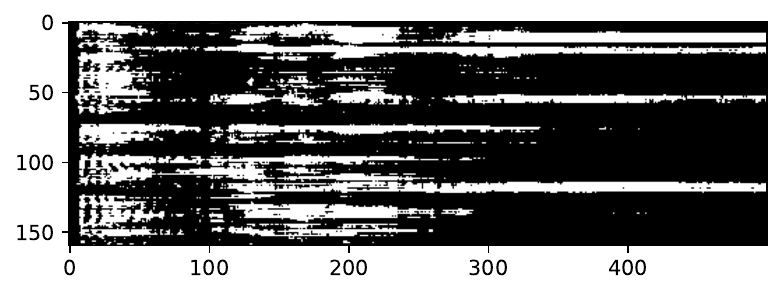} \\
    \end{tabular}}
  \centerline{
    \begin{tabular}{cccc}
	\includegraphics[width=0.235\textwidth]{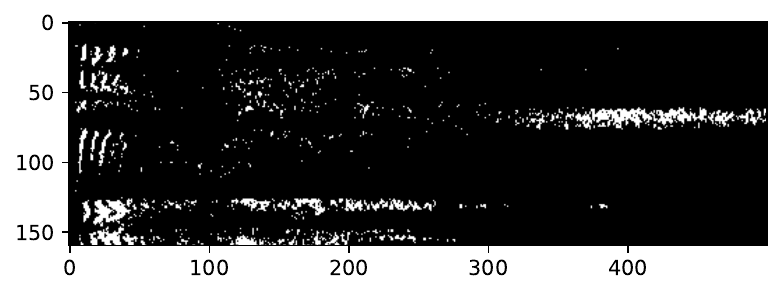} & 
	\includegraphics[width=0.235\textwidth]{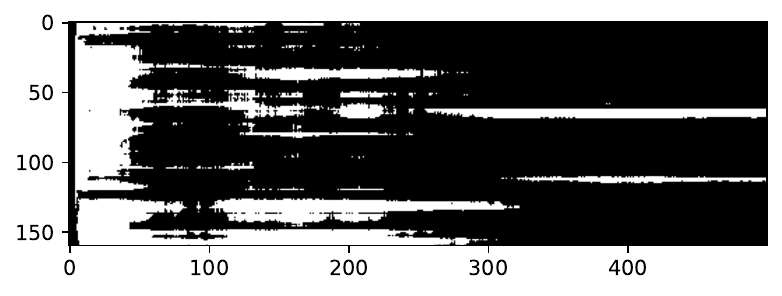} & 
	\includegraphics[width=0.235\textwidth]{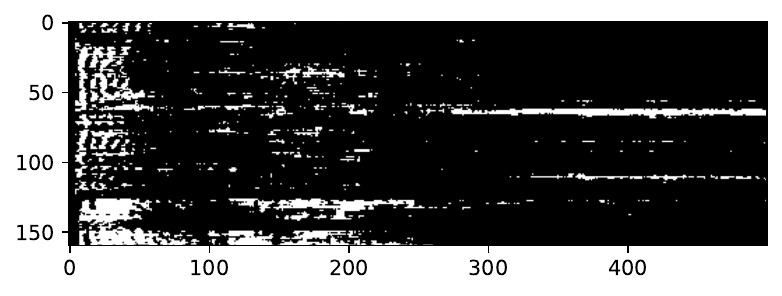} & 
	\includegraphics[width=0.235\textwidth]{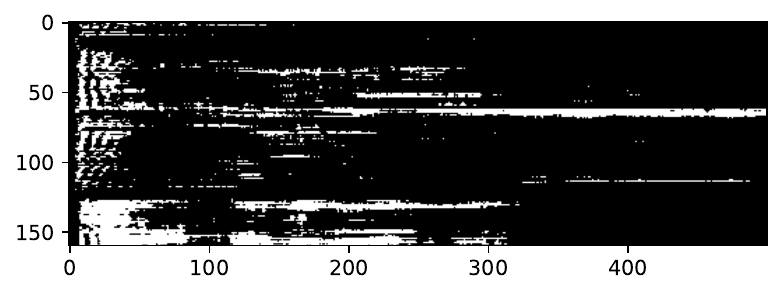} \\
    \end{tabular}}
  \centerline{
    \begin{tabular}{cccc}
	\includegraphics[width=0.235\textwidth]{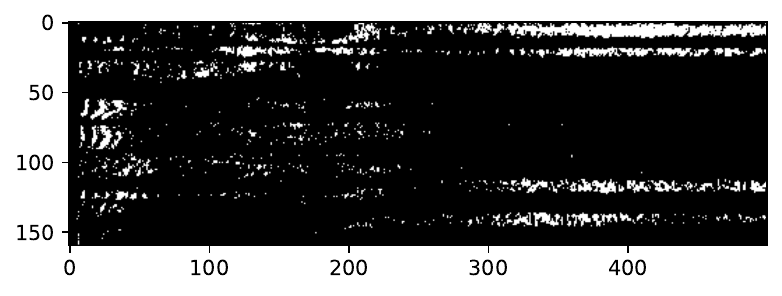} & 
	\includegraphics[width=0.235\textwidth]{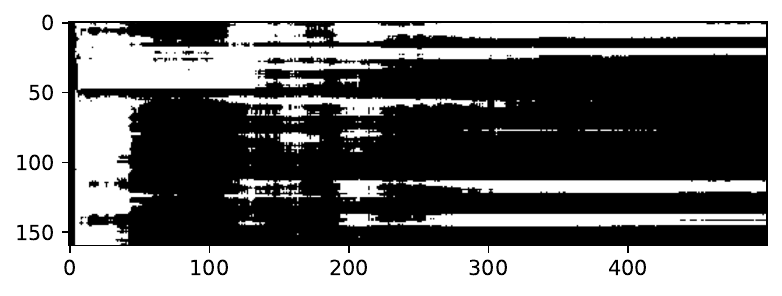} & 
	\includegraphics[width=0.235\textwidth]{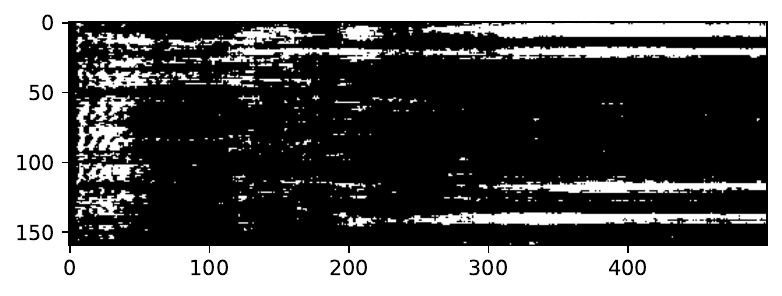} & 
	\includegraphics[width=0.235\textwidth]{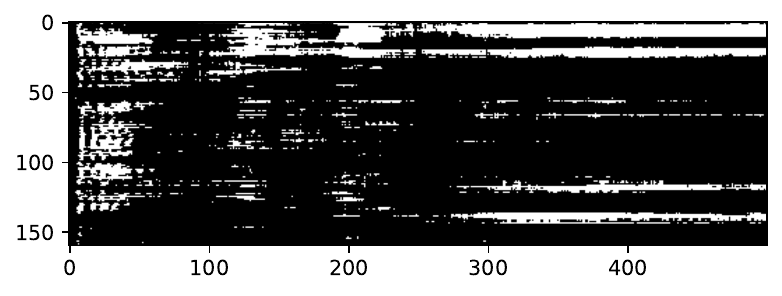} \\
    \end{tabular}}
    \caption{Mask reconstructions: (a) Original mask, (b) reconstruction using Backpropagation, and (c) Burstpropagation.}
  \label{f.reconstructions}
\end{figure*}

\subsection{Experimental Setup}
\label{ss.setup}

This work implements MBURST, a Burstpropagation-based multimodal architecture for speech enhancement. The architecture comprises two parallel networks, i.e., an audio and a visual channel, each containing two convolutional layers with $32$ channels, followed by flattening and a dense layer with $256$ units for embedding extraction. Finally, such embeddings are concatenated and used to feed the top layer, which is encharged to estimate the clean audio masks. It also implements two baselines for comparison purposes, i.e., a similar network using the same configuration but the backpropagation algorithm for learning, namely Multimodal and a unimodal version also trained using backpropagation that comprises the same architecture except by the visual channel

The convolutions consider kernels of size $3\times3$, paddings equal to $1$, and a stride of size $2$ for the visual and $1$ for the audio channels. This difference in the stride concerns the shape of the audio input, i.e., since the input audio comprises the current signal, composed of $500$ features, plus the seven prior frames, forming an $8\times500$ matrix, we opted to use stride $1$ to maintain this shape due to the reduced size. 

The experiments were conducted using the weighted binary cross entropy as the loss function since the rate of white pixels in the masks (relevant information) is several times smaller than black ones. The same justificative is considered for employing the F1-score as the evaluation metric. Further, the optimization of the parameters is conducted using Adam with a learning rate of $10^{-3}$ and weight decay of $10^{-6}$. Regarding Burstpropagation, the reference bursting probability follows the standard value adopted by the original work~\cite{payeur2021burst}, setting $\Bar{p}_{L}=0.2$. The dataset was randomly split into training ($80\%$), testing ($15\%$), and proxy ($5\%$) sets. The training is conducted for $3$ runs, considering different dataset splits, during $40$ epochs.

Finally, MBURST, as well as the baseline architectures, were implemented in Python using Pytorch framework~\cite{NEURIPS2019_9015}

\section{Experimental Results}
\label{s.experiments}

This section presents the results and discussion under the perspectives of clean audio signal mask reconstruction and energy efficiency.

\subsection{Clean Audio Signal Mask Reconstruction}
\label{ss.maskReconstruction}

Table~\ref{t.metrics} presents the average F1-score, accuracy, and the respective standard deviations, over the training and test sets for the proposed MBURST and the baseline architectures. In this scenario, one can notice that the Unimodel does not have the capacity to provide competitive results, which is expected since the visual context is essential to introduce context to the noisy signal and boost the reconstruction. On the other hand, both the Multimodal and MBURST approaches obtained similar results, with the Multimodal approach achieving slightly better values (less than $2\%$ on average). This can possibly be explained by the activation sparsity induced by the burst rate mechanism. Such results reinforce the relevance of visual context for clean speech reconstruction.

\begin{table}[!htb]
	\caption{F1-score and Accuracy concerning the Unimodal, Multimodal, and the proposed MBURST for clean audio mask reconstruction over the train and test sets.}
	\begin{center}
		\resizebox{\columnwidth}{!}{
			\begin{tabular}{cccccc}
				\toprule
				 Data set & Metric & Unimodal & Multimodal & MBURST \\
				\midrule
				\multirow{2}{*}{\textbf{Train}} & F1 & $0.677 \pm 0.000$ & $0.802 \pm 0.005$ & $0.782 \pm 0.001$ \\ 
				& Acc. & $82.919 \pm 0.043$ & $91.898 \pm 0.220$ & $90.602 \pm 0.067$ \\
				\midrule
				\multirow{2}{*}{\textbf{Test}} & F1 & $0.696 \pm 0.000$ &$0.790 \pm 0.003$ & $0.768 \pm 0.002$ \\ 
				& Acc. & $81.758 \pm 0.047$ & $90.236 \pm 0.164$ & $88.798 \pm 0.233$ \\
				\bottomrule
		\end{tabular}}
		\label{t.metrics}
	\end{center}
\end{table}

Figure~\ref{f.reconstructions} depicts some examples of the reconstructions themselves. Notice that such images were generated by collecting groups of $150$ randomly selected sequential samples from the proxy data set. In this context, one can observe that the masks reflect the results provided in Table~\ref{t.metrics}, i.e., Unimodal presents a poor mask reconstruction, while both the Multimodal and the MBURST provide more accurate and very similar results. Once again, the results suggest that a correlated visual context is essential to extract more relevant information from noisy audio signals and improve reconstruction quality.

\subsection{Energy Analysis}
\label{ss.energy}

\begin{figure}[!h]
  \centerline{
    \begin{tabular}{c}
	\includegraphics[width=0.75\columnwidth] {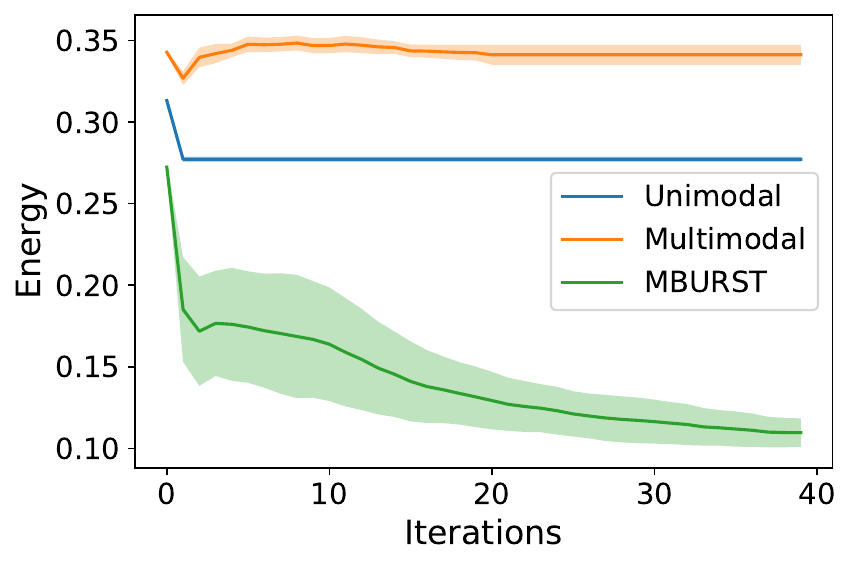} \\
        (a) 
    \end{tabular}}
  \centerline{
    \begin{tabular}{c}
        \includegraphics[width=0.75\columnwidth] {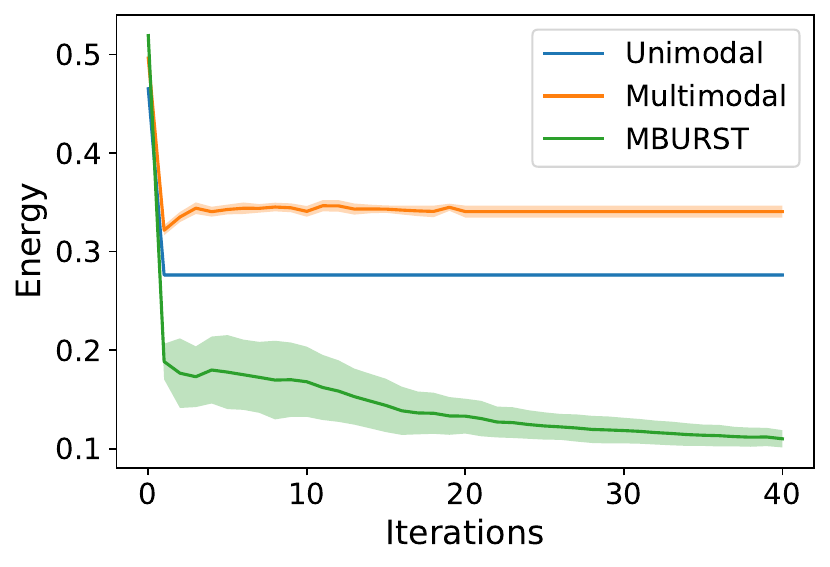}\\
        (b)
    \end{tabular}}
    \caption{Energy efficiency rate over (a) train and (b) test sets.}
  \label{f.energy}
\end{figure}

This work considers the energy efficiency in terms of neurons' activation rate, which shows itself as MBURST's major triumph. In this aspect, one can observe in Figure~\ref{f.energy} that the firing rate of MBURST neurons decreases dramatically to values below $0.2$ in the first iterations and strides towards values close to $0.1$ in subsequent iterations for both training and testing sets, showing an energy gain around $70\%$ and $65\%$ over the Multimodal and Unimodal approaches, respectively. Regarding the backpropagation-based approaches, the Unimodel performed better than the Multimodel, suggesting that the backpropagation was not robust enough to combine the noisy audio and visual information in an energy-efficiency fashion, thus implying the worst performance achieved by the Multimodel approach in this context.

Table~\ref{t.energy} analytically reinforces MBURST's robustness in the context of energy efficiency by presenting the area under the curve over the average energy rate measured during the training and testing procedures. In this context, MBURST shows itself $48\%$ and $58\%$, on average over train and test sets, more efficient than Unimodal and Multimodal, respectively.

\begin{table}[!htb]
	\caption{Area under the curve considering the average energy rate during training and testing for the Unimodal, Multimodal, and the proposed MBURST.}
	\begin{center}
		\resizebox{0.8\columnwidth}{!}{
			\begin{tabular}{ccccc}
				\toprule
				 Data set& Unimodal& Multimodal & MBURST \\
				\midrule
				\textbf{Train} &  $11.12$ &  $13.70$ & $5.63$ \\
				\midrule
				\textbf{Test} &  $11.51$ &   $14.14$ & $6.08$ \\
				\bottomrule
		\end{tabular}}
		\label{t.energy}
	\end{center}
\end{table}

\section{Conclusions}
\label{s.conclusions}

This paper proposed the MBURST, a Burstpropagation-based multimodal approach that implements a more biologically plausible solution for the task of AV speech enhancement. One can draw two main conclusions from experiments conducted over a Grid Corpus and CHiME3-based dataset. First, multimodal AV architectures are more efficient than a standard unimodal approach for clean audio mask reconstruction since correlated visual information provides contextual meaning to noisy audio. Second, MBURST can produce similar results to a traditional backpropagation multimodal architecture in the context of accuracy and reconstruction task metrics,  with a massive advantage regarding energy efficiency due to the burst-rate-based activation mechanism.

Regarding future work, we aim to extend the Burstpropagation concepts to recent studies comprising context-sensitive neocortical neurons~\cite{adeel2022context} and canonical cortical networks~\cite{passos2022canonical}.

\section*{Acknowledgments}
The authors are grateful to the Engineering and Physical Sciences Research Council (EPSRC) grant EP/T021063/1.

\bibliographystyle{IEEEtran}
\bibliography{references}
\end{document}